\pdfoutput=1
\documentclass[twocolumn,english,aps,superscriptaddress,prb]{revtex4-1}

\usepackage{babel}
\usepackage{amsmath}
\usepackage{amssymb}
\usepackage{graphicx}

\newcommand{\be}{\begin{equation}}
\newcommand{\ee}{\end{equation}}
\newcommand{\bea}{\begin{eqnarray}}
\newcommand{\eea}{\end{eqnarray}}

\begin{document}

\title{Comment on `Frustration and Multicriticality in the Antiferromagnetic Spin-1 Chain'}

\author{Natalia Chepiga}
\affiliation{Institute of Physics, Ecole Polytechnique F\'ed\'erale de Lausanne (EPFL), CH-1015 Lausanne, Switzerland}
\author{Ian Affleck}
\affiliation{Department of Physics and Astronomy, University of British Columbia, Vancouver, BC, Canada V6T 1Z1}
\author{Fr\'ed\'eric Mila}
\affiliation{Institute of Physics, Ecole Polytechnique F\'ed\'erale de Lausanne (EPFL), CH-1015 Lausanne, Switzerland}

\date{\today}
\begin{abstract} 
The phase diagram of the spin-1 chain with bilinear-biquadratic and next-nearest neighbor interactions, recently investigated by Pixley, Shashi and Nevidomskyy [Phys. Rev. B \textbf{90}, 214426 (2014)], has been revisited in the light of results we have recently obtained on a similar model. Combining extensive Density Matrix Renormalization Group (DMRG) simulations with conformal-field theory arguments, we confirm the presence of the three phases identified by Pixley et al, a Haldane phase, a next-nearest neighbor (NNN) Haldane phase, and a dimerized phase, but we come to significantly different conclusions regarding the nature of the phase transitions to the dimerized phase: i) We provide numerical evidence of a continuous Ising transition between the NNN-Haldane phase and the dimerized phase; ii) We show that the tri-critical end point, where the continuous transition between the Haldane phase and the dimerized phase turns into a first order transition, is distinct from the triple point where the three phases meet; iii) Finally, we demonstrate that the tri-critical end point is in the same Wess-Zumino-Witten (WZW) SU$(2)_2$ universality class as the continuous transition line that ends at this point.
\end{abstract}

\maketitle

%%%%%%%%%%%%%%%%%%%%%%%%%%%%%%%%%%%%% INTRODUCTION %%%%%%%%%%%%%%%%%%%%%%%%%%%%%%%%%%%%

\section{Motivation}
Two years ago, the phase diagram of the bilinear-biquadratic spin-1 chain with next-nearest neighbor interaction has been mapped out by Pixley, Shashi and Nevidomskyy \cite{nevidomskyy}. It consists of three phases, and the nature of the phase transitions has been determined using Density Matrix Renormalization Group (DMRG) and field-theory arguments. More recently, we have investigated a similar model in which the biquadratic interaction
is replaced by a three-site interaction that provides the appropriate generalization of the spin-1/2 Majumdar-Ghosh chain\cite{J1J2J3_letter}. 
Much to our surprise, while
the competing phases are the same as for the model with biquadratic interaction - Haldane, NNN-Haldane (called NNN-AKLT in Ref. [\onlinecite{nevidomskyy}]) and dimerized -  we came to significantly different conclusions regarding the transitions between them. The aim of this comment is to re-investigate the nature of the phase transitions in the model with biquadratic interactions along the lines of Ref. [\onlinecite{J1J2J3_letter}]. As we will see, this leads to a new phase diagram that turns out to be qualitatively similar to that of the model with three-site interactions.

\section{Phase Diagram}
\begin{figure}[h!]
\includegraphics[width=0.47\textwidth]{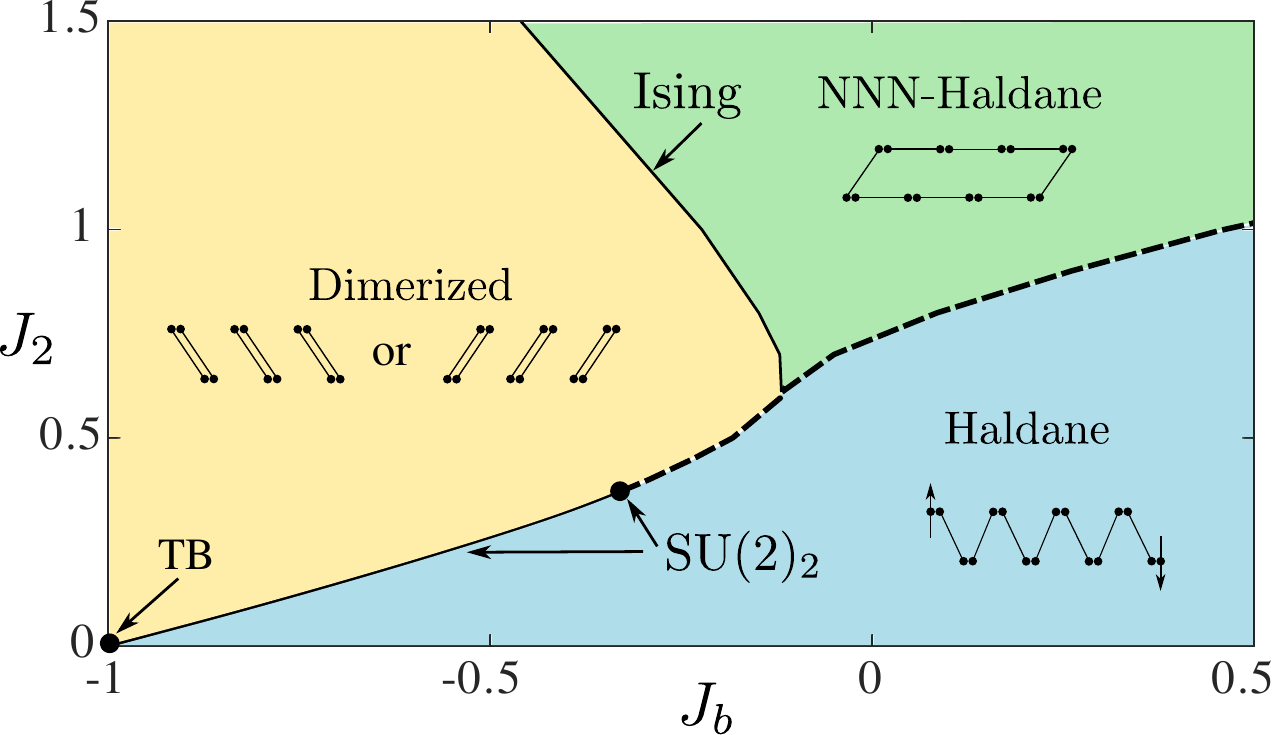}
\caption{(Color online) Phase diagram of the spin-1 chain with next-nearest neighbor coupling $J_2$ and biquadratic interaction $J_b$. The transition from the dimerized phase to the Haldane phase starts at the Takhtajan-Babudjian (TB) point, is continuous along the solid line, with central charge $c=3/2$, and first order along the dashed line. The transition from the NNN-Haldane phase to the dimerized phase is a continuous transition in the Ising universality class with central charge $c=1/2$. The transition between the Haldane phase and the NNN-Haldane phase is always first order.}
\label{fig:phase_diagram}
\end{figure}

The $J_1-J_2-J_b$ model is described by the Hamiltonian:
\begin{equation}
\label{eq:H}
H=\sum_{i}J_1{\bf S}_i\cdot {\bf S}_{i+1}+J_2{\bf S}_{i-1} \cdot {\bf S}_{i+1}+J_b({\bf S}_{i-1} \cdot {\bf S}_{i})^2,
\end{equation}
$J_1=1$ throughout the paper. In the convention of Ref-[\onlinecite{nevidomskyy}], $J_2=\alpha$ and $J_b=\beta$. Our main results are summarized in the phase diagram of Fig.\ref{fig:phase_diagram}. Each phase may be schematically illustrated  by valence bond pictures.

Below, with the help of extensive density matrix renormalization group (DMRG)\cite{dmrg1,dmrg2,dmrg3,dmrg4} calculations, we will demonstrate that:
{\it i)} The phase transition between the NNN-Haldane phase and the dimerized phase is continuous and in  the Ising universality class, and {\it not} first order;
{\it ii)} The continuous WZW SU$(2)_2$ transition starts at the Takhtajan-Babujian (TB) point and terminates at a tri-critical point that is {\it distinct} from the triple point;
{\it iii)} Beyond the tri-critical point, the phase transition between the Haldane phase and the dimerized phase is first order;
{\it iv)} The tri-critical point is in the same WZW SU$(2)_2$ universality class as the critical line that ends at that point, and {\it not} in the WZW SU$(2)_4$ universality class, as suggested in Ref. [\onlinecite{nevidomskyy}].

\section{Ising transition}

\begin{figure}[h!]
\includegraphics[width=0.47\textwidth]{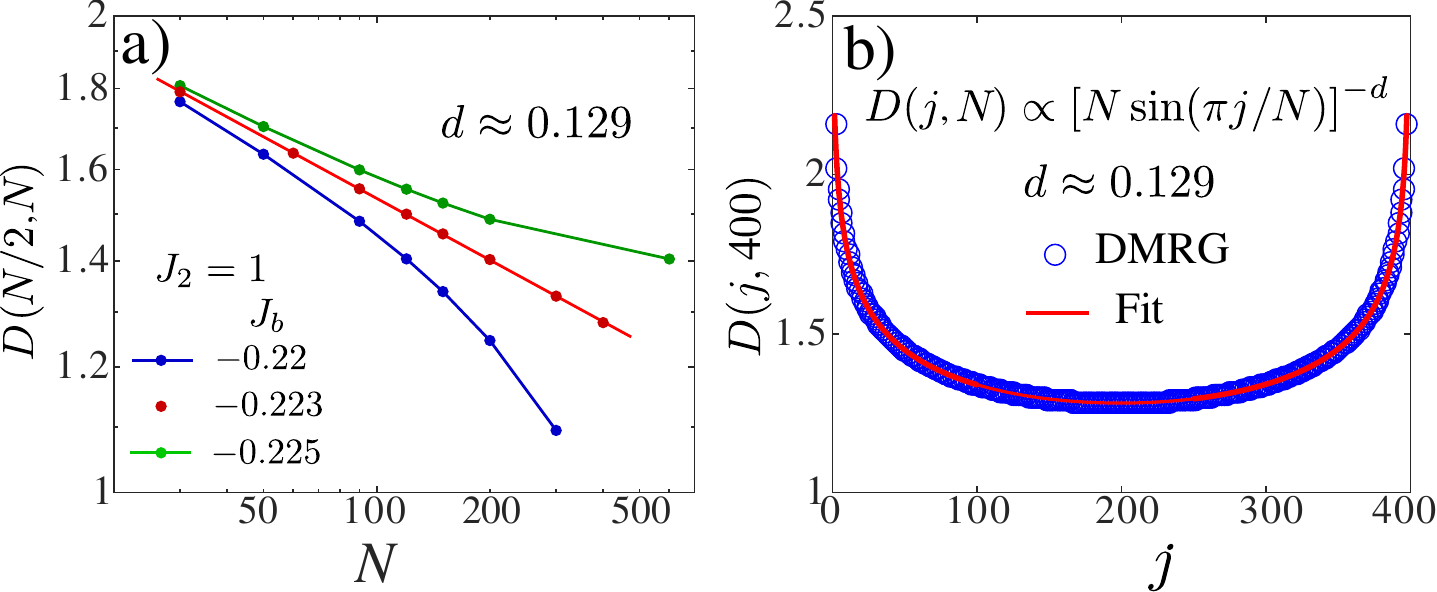}
\caption{(Color online) a) Log-log plot of the mid-chain dimerization in open chain as a function of the number of sites $N$ for $J_2=1$ and different values of $J_b$. The linear curve corresponds to the Ising critical point, and its slope to the critical exponent $d$. This leads to $J_{b}=-0.223$ and $d=0.129$, in good agreement with the prediction $1/8$ for Ising. b) Site dependence of $D(j,N)$ at the critical point fitted to $1/\left[N\sin (\pi j/N)\right]^d$. This leads to an exponent $d=0.128$, again close to Ising prediction $1/8$.}
\label{fig:cri_exp_ising}
\end{figure}

Let us first consider the transition between the NNN-Haldane phase and the dimerized phase.
We define the local dimerization as $D(j,N)=|\langle\vec{S}_{j}\cdot\vec{S}_{j+1} \rangle-\langle \vec{S}_{j-1}\cdot\vec{S}_{j}  \rangle|$, where $j$ is the spin index and $N$ is the total number of spins.
In order to locate the phase boundaries, we look at the mid-chain dimerization $D(N/2,N)$ around the transition as a function of system size $N$. In the NNN-Haldane phase, the dimerization vanishes with the system size, while in the dimerized phase it stays finite. 
In finite-size chains, we found that the dimerization increases continuously from NNN-Haldane phase to the dimerized phase, in agreement with the numerical results of Ref. [\onlinecite{nevidomskyy}]. 
The separatrix in a log-log plot corresponds to the phase transition, and its slope is equal to the critical exponent (see Fig.\ref{fig:cri_exp_ising}). Since open boundaries favor dimerization, they correspond to non-zero boundary magnetic field in the Ising model. From boundary conformal field theory (CFT), the magnetization at the critical point is expected to decay away from the boundary as [\onlinecite{cardy91}] $\langle\sigma(x)\rangle\propto 1/x^{1/8}$. Moreover, for a finite system $\langle\sigma(x)\rangle\propto 1/[(N/\pi)\sin(\pi x/N)]^{1/8}$. Identifying the local dimerization with $\sigma(x)$, one gets $D(j,N)\propto 1/[N\sin(\pi j/N)]^{1/8}$ and in particular $D(N/2,N)\propto1/N^{1/8}$. The  critical exponent obtained numerically $d\approx0.129$ is in good agreement with the Ising prediction.

\begin{figure}[h!]
\includegraphics[width=0.45\textwidth]{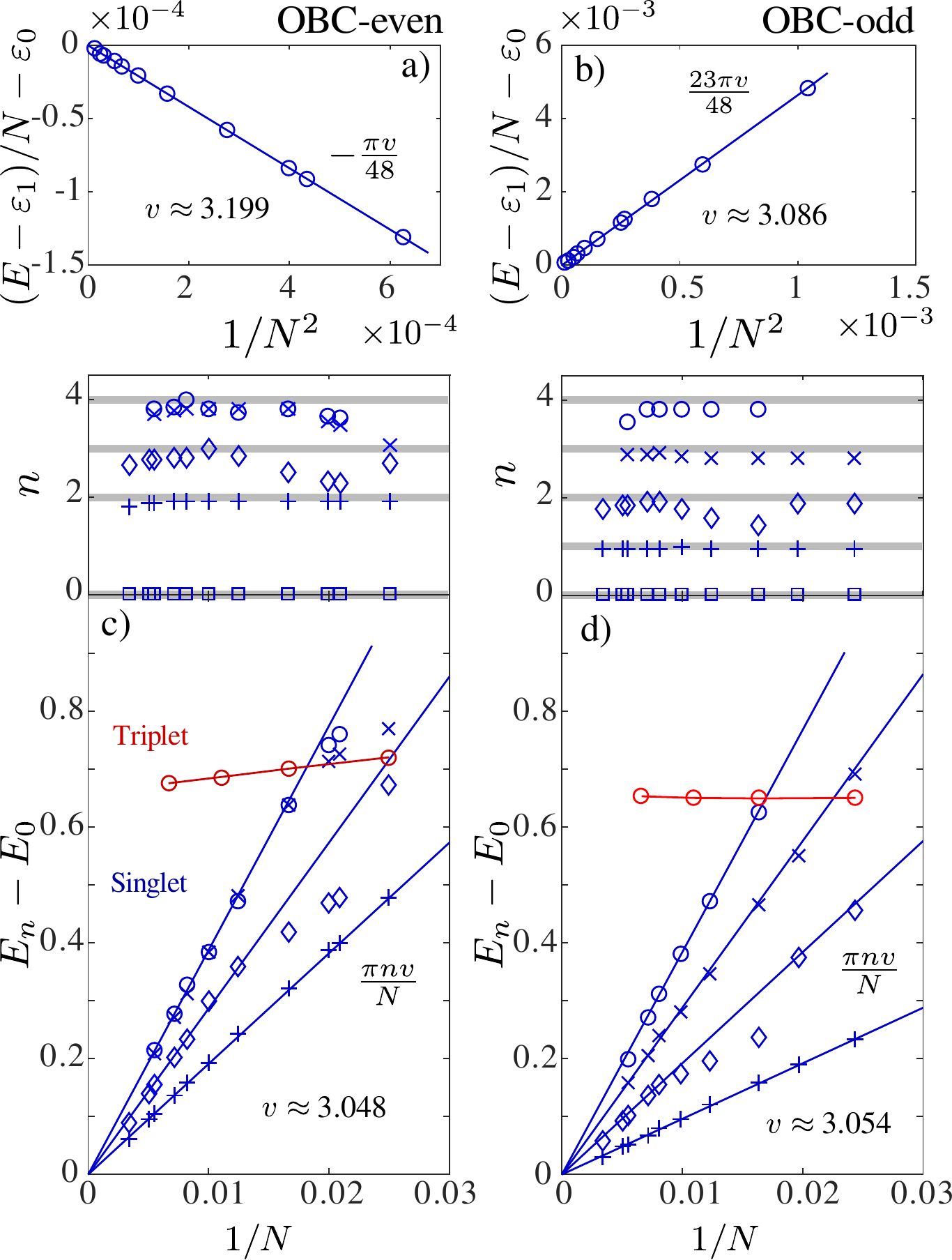}
\caption{(Color online) Ground-state and excitation energy at $J_2=1$ and $J_b=-0.223$, a point that belongs to the Ising critical line. a) and b) Linear scaling of the ground-state energy per site in an open chain with $1/N^2$ after subtracting the $\varepsilon_0$ and $\varepsilon_1$ terms for even and odd number of sites. c) and d) Energy gaps in the singlet and triplet sectors for OBC as a  function of $1/N$ for even and odd number of sites. The slope of the singlet gap gives access to the value of the velocity. Insets: Conformal towers. Grey lines show Ising conformal towers for $I$ ($N$ even) and for $\epsilon$ ($N$ odd). Blue symbols correspond to DMRG data.}
\label{fig:conf_tow_ising}
\end{figure}

We identify open boundary conditions in our model with $\uparrow ,\uparrow$ boundary conditions in the Ising model for $N$ even and with $\uparrow ,\downarrow$ boundary conditions for $N$ odd, where the arrows refer to the directions of boundary magnetic fields in the Ising model\cite{J1J2J3_letter}. Then according to conformal field theory (CFT), the ground state energy in an open Ising chain scales with the system size $N$ as $E=\varepsilon_0N+\varepsilon_1-\pi v/(48N)$ for $N$ even and $E=\varepsilon_0N+\varepsilon_1+\pi v/(23N)$ for $N$ odd. The scaling of the DMRG data for the ground-state energy are presented in Fig.\ref{fig:conf_tow_ising}a-b).

We have calculated the lowest four excited state energies for both parities of $N$ in the singlet sector as well as the triplet gap; see Fig.\ref{fig:conf_tow_ising}c-d). The excitation energies in singlet sector reveal the expected Ising conformal tower of $I$ for $N$ even and of $\epsilon$ for $N$ odd [\onlinecite{cardy89}]. This definitely
establishes that the transition is continuous and in the Ising universality class. 

By contrast, the singlet-triplet gap remains finite. 
In Ref. [\onlinecite{nevidomskyy}], the authors came to the same conclusion regarding the singlet-triplet gap. However, they did not investigate the singlet
sector. So they came to the conclusion that there is no gap closing at the transition, and accordingly that the transition must be first order. 

\section{Transition between Haldane and dimerized phases}

\begin{figure}[h!]
\includegraphics[width=0.47\textwidth]{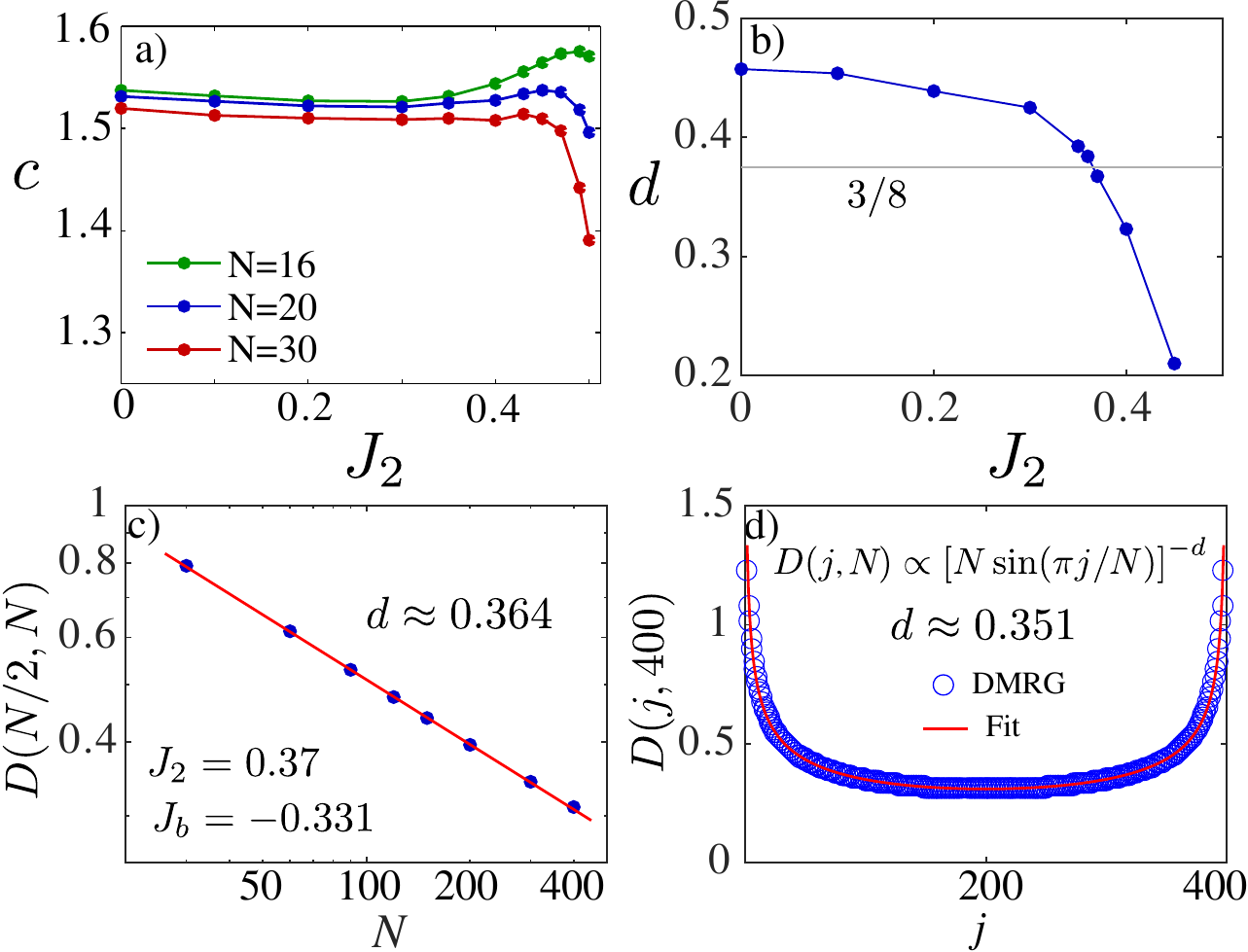}
\caption{(Color online) a) Central charge along the critical line as determined from fitting the entanglement entropy of periodic chains with the Calabrese-Cardy formula\cite{CalabreseCardy}. b) Apparent critical exponent along the SU$(2)_2$ critical line as a function of $J_2$ from the slope of the log-log plot $D(N/2,N)$ as a function of $N$ for the values $J_b$ for which it is linear. The grey line is the theoretical value of the exponent, $3/8$. c) Log-log plot of the mid-chain dimerization as a function of the number of sites $N$ at the critical point $J_2=0.37$ and $J_b=-0.331$. The slope corresponds to the critical exponent $d=0.364$, in good agreement with $3/8$ for WZW SU$(2)_2$. d) Site dependence of $D(j,N)$ at the critical point fitted to $1/\left[N\sin (\pi j/N)\right]^d$. This leads to an exponent $d=0.351$, again close to WZW SU$(2)_2$ prediction $3/8$.}
\label{fig:cri_exp_su2}
\end{figure}

As mentioned above, the transition between the Haldane phase and the dimerized phase starts at the TB point, where it is continuous in the WZW SU$(2)_2$ universality class, and terminates at the tri-critical point, where the transition becomes first order. The SU$(2)_2$ phase transition is characterized by the critical exponent $d=3/8$ and the central charge $c=3/2$. 

In Ref. [\onlinecite{nevidomskyy}], the main argument in favor of the WZW SU$(2)_4$ universality class at the end point was based on the central charge. It was extracted from the scaling of the entanglement entropy $S_N(n)$ with block size $n$ in open chains of size $N$ according to the Calabrese-Cardy formula\cite{CalabreseCardy}:
\begin{equation}
S(n)=\frac{c}{6}\ln\left[\frac{2N}\pi \sin\left(\frac{\pi n}{n}\right)\right]+s_1+\log g. 
\label{eq:calabrese_cardy}
\end{equation}
Using an open chain with $N=90$ sites, the authors of Ref. [\onlinecite{nevidomskyy}] came to the conclusion that the central charge is around $c=2$. 

Since the finite-size effects for open systems are usually quite strong for the extraction of the central charge, we have revisited this conclusion using periodic systems. In Fig.\ref{fig:cri_exp_su2}a), we present results for the central charge extracted from fits of the entanglement entropy to the Calabrese-Cardy formula for {\it periodic} systems:
\begin{equation}
S(n)=\frac{c}{3}\ln\left[\frac{N}\pi \sin\left(\frac{\pi n}{n}\right)\right]+s_1 
\label{eq:calabrese_cardy}
\end{equation}
The results for $N=16$, $20$ and $30$ sites are shown. The central charge extracted from periodic chains has very small finite-size dependence, and it is clear that it never exceeds significantly the value $c=3/2$. This implies that the end point is in the WZW SU$(2)_2$ universality class. To recover this result
with open boundary conditions, one should presumably use systems with much more than $90$ sites.

\begin{figure}[h!]
\includegraphics[width=0.45\textwidth]{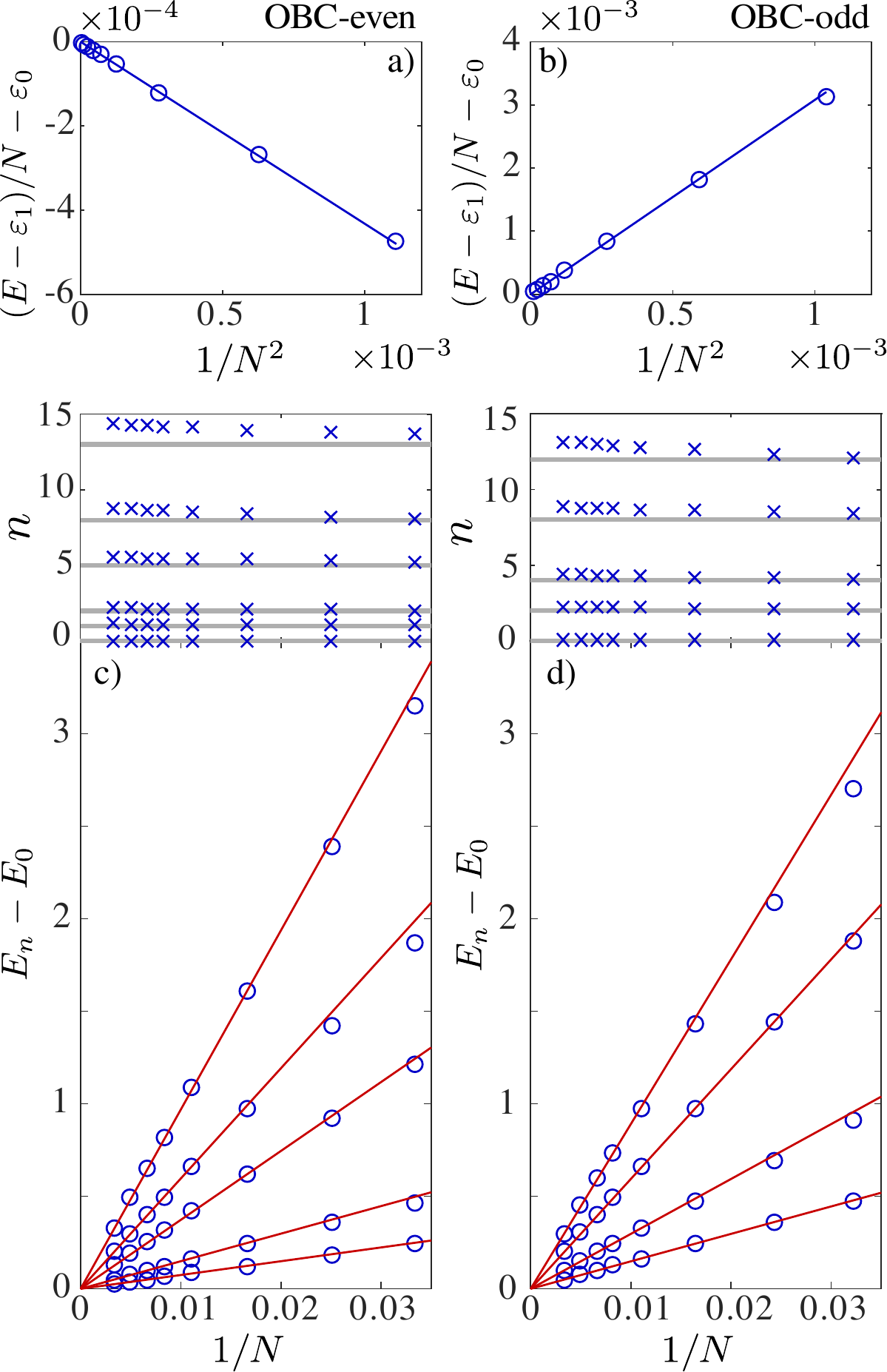}
\caption{(Color online) Ground-state and excitation energy at $J_2=0.37$ and $J_b=-0.331$, a point that belongs to the critical line between the Haldane and the Dimerized phases. a) and b) Linear scaling of the ground-state energy per site in an open chain with $1/N^2$ after subtracting the $\varepsilon_0$ and $\varepsilon_1$ terms for even and odd number of sites. c) and d) Energy gap between the ground state and the lowest energy states in different sectors of $S_z^\mathrm{tot}=0,1,...,5$ as a function of $1/N$ for even and odd number of sites. Insets: Conformal towers. Blue symbols correspond to DMRG data. Red lines are the expected conformal towers \cite{J1J2J3_letter}, with a velocity defined by the finite-size scaling of the ground state energy for $N$ even.}
\label{fig:conf_tow_su2}
\end{figure}

We now confirm these results by calculating the critical exponent and the conformal towers. The tri-critical point is characterized by the absence of logarithmic corrections. This is the only point along the critical line where the critical exponents can be accurately extracted from finite sizes. We again look for the separatrix in the scaling of the mid-chain dimerization in order to locate the critical line, as described in the previous section. The slope gives an apparent critical exponent, presented in Fig.\ref{fig:cri_exp_su2}c). The point at which the slope is the closest to the predicted value $3/8$ (Fig.\ref{fig:cri_exp_su2}a) is identified with the end point. The critical exponent obtained at the end point from a scaling analysis of the dimerization $D(j,N)$ with the spin position $j$ for a fixed chain length $N$ is also in good agreement with the prediction $d=3/8$; see Fig.\ref{fig:cri_exp_su2}b). The position of the end point deduced from this analysis
is $J_2=0.37$, $J_b=-0.331$, well separated from the triple point. 

In Ref. [\onlinecite{nevidomskyy}] it was suggested that these two points coincide.  While the estimate of the triple point $0.47<J_2<0.55$ and $-0.2<J_b<-0.15$ reported in Ref. [\onlinecite{nevidomskyy}] is consistent with our results, we think that the two points do not coincide, and that the tri-critical point lies clearly
outside this interval.

\begin{figure}[h!]
\includegraphics[width=0.47\textwidth]{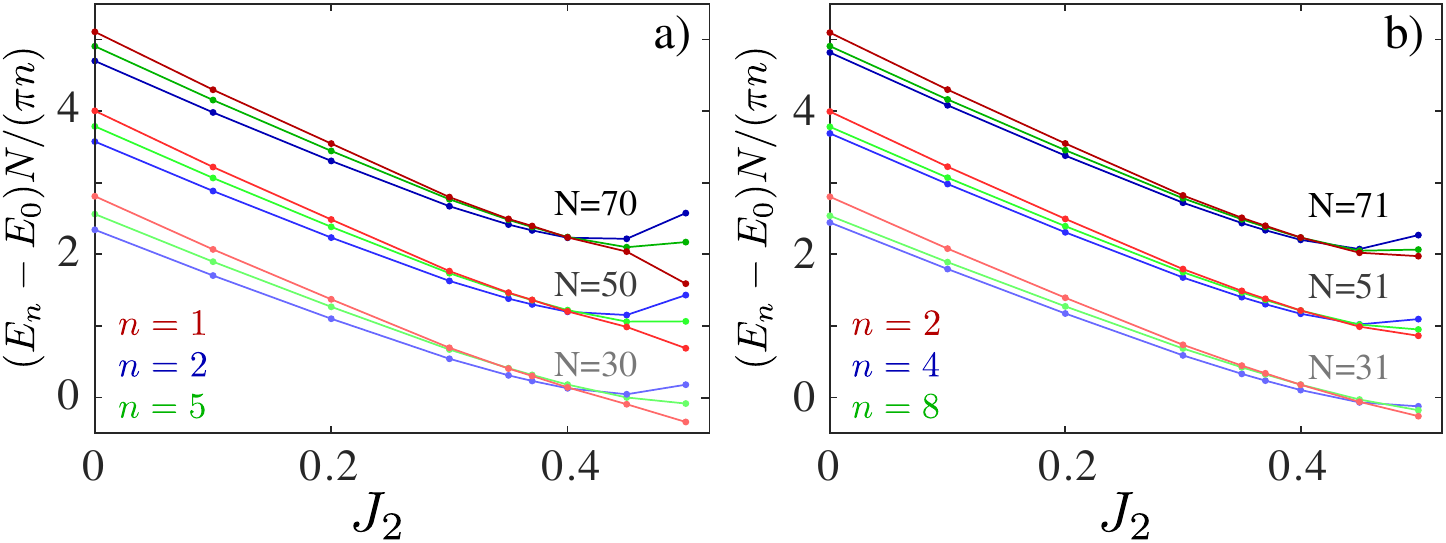}
\caption{(Color online) Velocity along the critical line between the Haldane phase and the dimerized phase extracted from the gap between various energy levels and the ground state. For clarity, results for $N=50$, $51$ and $N=70$, $31$ are shifted vertically by 1 and 2 respectively.}
\label{fig:velocities_su2}
\end{figure}

The ground-state energies for even and odd number of sites are expected to scale according to\cite{J1J2J3_letter}:
\begin{equation}
E_\mathrm{even}=\varepsilon_0 N+\varepsilon_1-\frac{\pi v}{16N}, \ \ E_\mathrm{odd}=\varepsilon_0 N+\varepsilon_1+\frac{7\pi v}{16N}.
\end{equation}

In order to build the conformal tower at the end point $J_2=0.37$ and $J_b=-0.331$, we calculate the gap between the ground-state energy and the lowest energies in different sectors of $S^z_\mathrm{tot}$. The gap scales linearly with $1/N$, and the slope gives access to the velocity. In a chain with an even number of sites, the ground state is a singlet and the first excited states is a triplet, while in a chain with an odd number of sites, the ground state is in the triplet sector. The DMRG data on the scaling are presented in Fig.\ref{fig:conf_tow_su2}.

In order to prove that the point $J_2=0.37$ and $J_b=-0.331$ is indeed the end point, we checked that the logarithmic corrections destroy the conformal towers away from this point \cite{J1J2J3_letter}. In order to do so, we have calculated the velocities by performing a linear fit of the gap for the first three levels in each tower. The conformal towers are reconstructed only when all velocities take the same values. Otherwise the structure is perturbed. Fig.\ref{fig:velocities_su2} provides examples of velocities extracted along the critical line for different sizes. The crossing points around $J_2=0.37$ are compatible with our determination of the tri-critical point.

\section{Conclusion}

Extensive DMRG calculations coupled to CFT arguments have revealed significant differences with the original phase diagram of Ref. [\onlinecite{nevidomskyy}] regarding the nature of the phase transitions: i) The phase transition between NNN-Haldane phase and dimerized phase turns out to be continuous, and in the Ising universality class; ii)  The tri-critical point at which the continuous WZW SU$(2)_2$ transition turns into a first order occurs below the triple point;
iii) This tri-critical point is in the same WZW SU$(2)_2$ universality class as the critical line that ends at this point.

The similarities of the phase diagrams for biquadratic and three-site interactions suggest that their main features are generic for the spontaneous dimerization transitions of spin-1 chains.

{\it Acknowledgments:} This work has been supported by the Swiss National Science Foundation and by NSERC Discovery Grant 36318-2009 and CIFAR (IA). 

\bibliographystyle{apsrev4-1}
\bibliography{bibliography}
%\begin{thebibliography}{100}
%\bibliography{bibliography}
%\bibitem{WZW} The end of the critical line does {\it not} correspond to a $k>2$ WZW model, but simply to the point where $k=2$ WZW behavior is most clearly  observed,  due to the vanishing of the marginally irrelevant coupling constant there.  We disagree in this regard with Ref.\onlinecite{nevidomskyy}.
%\end{thebibliography}
\end{document}